\documentclass[11pt]{article}

\usepackage[iso-8859-7]{inputenc}
\usepackage{epsfig}
\usepackage{graphicx}
\usepackage{epstopdf}
\usepackage{amsfonts}
\usepackage{amssymb}
\usepackage{amsthm}
\usepackage{amsmath}
\usepackage{multirow}

\newcommand{\newc}{\newcommand}
\newc{\ra}{\rightarrow}
\newc{\lra}{\leftrightarrow}
\newc{\be}{\begin{equation}}
\newc{\ee}{\end{equation}}
\newc{\ba}{\begin{eqnarray}}
\newc{\ea}{\end{eqnarray}}
\newc{\ov}{\overline}
\newc{\pa}{\partial}
\newc{\D}{\Delta}

\newc{\nn}{\nonumber}

\begin{document}
\thispagestyle{empty}

\hfill CERN-PH-TH/2014-073
\vskip 2truecm
\vspace*{3cm}
\begin{center}
{ {\bf {\cal F}-GUTs  with Mordell-Weil U(1)'s}}\\
\vspace*{1cm}
{\bf
I. Antoniadis$^{1,\,\flat }$,   G.K. Leontaris$^{2}$}\\
\vspace{4mm}
$^1$  Department of Physics, CERN Theory Division, \\
CH-1211, Geneva 23, Switzerland\vspace{1mm}\\
$^2$ Physics Department, Theory Division, Ioannina University, \\
GR-45110 Ioannina, Greece\vspace{1mm}\\
\end{center}

\vspace*{1cm}
\begin{center}
{\bf Abstract}
\end{center}
\noindent

In this note we study the constraints on F-theory GUTs with extra $U(1)$'s in the context of elliptic fibrations
with rational sections. We consider the simplest case 
of one  abelian factor (Mordell-Weil rank one)  and investigate the conditions that are induced on the coefficients 
of   its Tate form.
 Converting the equation representing the generic hypersurface  $P_{112}$
 to this Tate's form we find that the presence of a U(1), already in this local description, is consistent with
  the exceptional ${\cal E}_6$ and  ${\cal E}_7$ non-abelian singularities.  We briefly comment on a viable
${\cal E}_6\times U(1)$ effective  F-theory model.

\vfill
$^{\flat}$ {\it On leave from CPHT (UMR CNRS 7644) Ecole Polytechnique, F-91128 Palaiseau, France.}

\newpage

\section{Introduction}

It has been by now widely accepted that additional $U(1)$ or discrete symmetries constitute an important
ingredient  in GUT model building. Such symmetries are useful to prevent  dangerous superpotential couplings
of the effective field theory model, in particular those inducing proton decay operators and lepton number
violating interactions at unacceptable rates.  Model building in the context
of string theory has shown that such symmetries are naturally incorporated in the emerging
effective field theory model. In the context of
 F-theory~\cite{Vafa:1996xn}
  in particular,  the last few years several GUT symmetries have been analysed with the presence of
 additional $U(1)$ factors~\cite{Beasley:2008dc}
\footnote{For an incomplete list see\cite{Beasley:2008kw}-\cite{King:2010mq},
the reviews~\cite{Heckman:2010bq,Weigand:2010wm,Leontaris:2012mh,Maharana:2012tu} and references
 therein.}.

In  F-theory models the non-abelian part of the gauge group is determined by specific
geometric singularities of the internal manifold.   The internal space is
 an elliptically fibred Calabi-Yau (CY) fourfold $Y_4$,  over a three-fold base $B_3$.
The fibration is  determined by the Weierstra\ss\,   model
\be
y^2 = x^3+f(\xi)\,xz^4+g(\xi)\,z^6\label{Weixyz}
\ee
where the base of the fibration corresponds to the point of the torus $z\to 0$ and as such it defines a
zero section at $[x:y:z]=[t^2:t^3:0]$. For particularly restricted  $f,g$ functions  the fiber degenerates
 over certain points of the base. The  non-abelian  singularities of the fiber
  are well known and have been systematically classified with respect to the vanishing order of the functions $f,g$
  and the roots of the discriminant of (\ref{Weixyz}),  by Kodaira~\cite{Kodaira:1963}. An equivalent description useful for
 local  model building is also given by Tate~\cite{Tate1975,Bershadsky:1996nh}.  There are $U(1)$ symmetries however which do not
  emerge from a non-abelian singularity and as such they do not fall into the category of a Cartan subalgebra.
  There is no classification for such U(1) symmetries analogous to the non-abelian case  and up to now they
   have not been fully explored.    Abelian factors correspond to extra rational sections and as such
 they imply additional restrictions  on the form of the functions $f,g$. Because
  sections are given in terms of divisors whose intersection points with the fiber should be distinct and not
  identifiable by any monodromy action, this can occur only for rational intersection points. Therefore,
 for such  points  of an elliptic curve fibered over $B_3$, their corresponding degree line bundle has a section that vanishes
  at these points.

  It is known  that rational points on elliptic curves constitute a group, the so called Mordell-Weil group.
   The Mordell Weil group is finitely generated in the sense that there exists a finite basis
    which generates all its elements~\cite{Silverman}.  A finitely generated  group can be written as
    \[ Z\oplus Z\oplus \cdots \oplus Z\oplus {\cal G}\]
    where    ${\cal G}$ is the torsion subgroup, which in principle could be  a source for useful discrete symmetries
  in the effective Lagrangian.
 Recent developments in F-theory have analysed some properties of the latter and its implications on effective field
 theory models. The rank of the abelian group is the rank of the Mordell-Weil group~\cite{Morrison:1996pp,Morrison:2012ei},
  however, the latter in not known. Up to now, studies  with one, two and three extra sections have appeared
  and some general implications on the low energy models have been accounted for~\cite{Cvetic:2012xn}-\cite{Krippendorf:2014xba}.
  
  In this note we argue that the appearance of extra sections has significant implications on the engineering of non-abelian gauge symmetries based on the local Tate form of the model.
  In particular, in the case of local constructions  based on the simple Tate's algorithm 
   the rational sections impose  certain restrictions on the defining equation. When the
  latter is converted to the familiar local Tate's form, in order to meet the requirements of the 
  extra rational section,  certain relations among the Tate's form coefficients occur.
 We will see that such constraints make impossible the appearance of familiar groups such as SU(5) in the local Tate form.
 To our knowledge, this issue has not been observed, and 
 it might  constitute another obstruction on the validity of simple Tate's algorithm similar to 
     those observed in reference~\cite{Katz:2011qp}. 
 Such  obstructions can be evaded in more general 
  models based on the `top' constructions of toric geometry~\cite{Bouchard:2003bu}. Using the latter techniques,
      $SU(5)$ models with several Mordell-Weil $U(1)$'s have been built~\cite{Cvetic:2012xn}-\cite{Krippendorf:2014xba}.
    However,  in this note we show that  in the context of the
   familiar local Tate's algorithm, viable effective models based on the exceptional singularities can be still easily
    accommodated. 
      
Therefore, it is the purpose of this note to examine the aforementioned 
 constraints and discuss the implications in the effective theory.
As a ``test ground'', we consider in particular the simplest case of two sections, i.e., one extra section in addition to the universal
one and since abelian factors are related to extra sections, this means that the GUT symmetry will be supported by
an extra $U(1)$.  Given the existence of one extra section, we re-derive the constraints on the Weierstra\ss\, model
 written in  Tate's form. Investigating the relations of the coefficients we find that there are basically
 two viable GUT symmetries, namely $E_6$ and $E_7$  supplemented by the extra abelian factor.  We briefly
 discuss the spectrum of the model $E_6\times U(1)$.

\section{Case of two rational points}

To set the stage, we recapitulate in this section some relevant results derived in~\cite{Morrison:2012ei}.
  In fact, we  re-consider 
thoroughly the derivation of the Weierstra\ss\, equation from the $P_{(1,1,2)}$ fibration with two rational sections.
As a result, in the process of converting the initial form   we find a second solution
which is distinct from the first one  with respect to the signs of the coefficients in  Tate's model.

 We consider  an elliptic curve ${\cal E}$  over a field ${\cal K}$, a point $P$ associated to the holomorphic (zero) section,
a rational point $Q$, and denote ${\cal M}={\cal O}(P+Q)$ the corresponding line bundle of degree 2. From the Riemann-Roch
theorem for genus one curves, we know that the number of global sections of a line bundle ${\cal M}$ is equal to its degree, $h^0({\cal M})=d$.
Because in our case $d=2$,  the group  $H^0({\cal M})$  must have two sections which we call them $u$ and $v$ with weights equal to 1.
 Considering now $H^0(2{\cal M})$, it can be seen that a new section $w$ with weight 2 is required, so that the
 three weights are $[u,v,w]=[1,1,2]$.
 Further, from $u,v,w$ one can form six  sections of degree 6 which match exactly the number of independent sections of
$H^0(3{\cal M})$, while all possible sections corresponding to $H^0(4{\cal M})$ that can be constructed are nine, exceeding
the independent ones by one.  Hence  there has to be a constraint among them which defines a hyper-surface in the weighted
projective space $P_{(1,1,2)}$ given by  the equation which relates them
\be
w^2+a_0u^2w+a_1uvw+a_2v^2w=b_0u^4+b_1u^3v+b_2u^2v^2+b_3uv^3+b_4 v^4
\label{hysu}
\ee
with $a_i, b_j$ coefficients in ${\cal K}$.

 One of the sections corresponds to the universal one so it vanishes  at the two points $P,Q$. We can take this to be the $u$
 section and therefore the equation (\ref{hysu}) at these points becomes
\be
w^2+a_2v^2w=b_4 v^4
\label{hysu0}
\ee
The roots of the equation correspond to the points $P,Q$ and since these are rational points the equation should split
 in two  factors, with all coefficients in the field ${\cal K}$.  To avoid square roots
we may redefine $\tilde w=w+\zeta v^2,  \tilde a_2^2=a_2^2+4 b_4$ with $2 \zeta=a_2-\tilde a_2$ and write this equation as
$\tilde w^2+\tilde a_2\tilde w v^2=0$.  Renaming  $\tilde w\to w$  for simplicity, we get
\[w(w+a_2v^2)=0\]
 whose roots are the points $P,Q$
 \[  [u:v:w]=[0:1:0]\; {\rm and }\; [u:v:w]=[0:1:-a_2]\]
  With this redefinition, we can eliminate the term
 $b_4v^4$ in the original equation (\ref{hysu}), while similar reasoning allows us to set $a_0=a_1=0$.
Under the aforementioned circumstances
the original equation reads~\footnote{Notice that the singularity is resolved by blowing up $w\to s w$ and $u\to s u$ so that
\[sw^2+a_2v^2w=u(b_0u^3s^3+b_1u^2s^2v+b_2usv^2+b_3v^3)\]}
\be
w^2+a_2v^2w=u(b_0u^3+b_1u^2v+b_2uv^2+b_3v^3)
\label{hysus}
\ee
To recover the Weierstra\ss\, form  with  global section  associated  to  $P$,
one has to find sections $H^0(k P)$. Since from group structure this is  $H^0(k M-k Q)$
one has to look for $H^0(kM)$ vanishing $k$ times at $Q$.

Starting with $k=1$, we have already assumed that the section $u$ vanishes at $P,Q$ and thus one can  set $u:=z$.
 For $k=2$ one section is $u^2$ while the other must be a linear combination of all possible degree-2 sections. Let
\[ w=\gamma  u^2+\beta  u v+\alpha  v^2\]
Substituting  in equation (\ref{hysu}) while organising in powers of $u$, we get
\[(\beta ^2+\gamma  (2 \alpha +a_2)-b_2)u^2+(\beta  (2 \alpha +a_2)-b_3)u+\alpha  (\alpha +a_2)\]
The vanishing of the  coefficients of zeroth and first order powers in $u$   above, gives the solutions

\[\alpha \,=\, -a_2,\beta \,=\, -\frac{b_3}{a_2}\]
and
\[\alpha \,=\, 0,\beta \,=\, \frac{b_3}{a_2}\]

Therefore, (setting $\gamma =0$ since section $u^2$ has already been included) we can have two
possible forms of the section $x$ given by
\be
\begin{split}
x&=b_3 uv +a_2 w+a_2^2 v^2\\
x&=b_3 u v -a_2 w
\end{split}
\ee

To find $y$ we examine $H^0(3 M)$. In general we expect another combination of the form
\[ w\,=\, \mu  u^2+\lambda  u v+\kappa  v^2\]
We substitute as before, and demand vanishing of the coefficients up to second order in $u$:
\[ \kappa  (a_2+\kappa )=0,\;\lambda  (a_2+2 \kappa )-b_3=0,\;\mu  (a_2+2 \kappa )-b_2+\lambda ^2=0 \]
Again, we obtain two distinct solutions which imply two forms of $y$:
\be
\begin{split}
y&=a_2^3 v^3+a_2^2 v w+a_2b_2 u^2 v-\frac{b_3^2 u^2 v}{a_2}+a_2   b_3 u v^2\\
   y&=a_2^2 v w-a_2 b_2 u^2 v+\frac{b_3^2 u^2 v}{a_2}-a_2b_3 u
      v^2
\end{split}
\ee

To recover the Weierstra\ss\, form of the original equation, we must invert the equations of $x(u,v,w),  y(u,v,w)$
and substitute them into the original equation.  On can observe that both sets of $x,y$ solutions leads to the same Weierstra\ss\,
form. For the first solution
\be
\begin{split}
v&=\frac{a_2 y}{a_2^2 \left(b_2 u^2+x\right)-b_3^2 u^2}
  \\
  w&=-\frac{a_2^3 y^2}{\left(b_3^2 u^2-a_2^2 \left(b_2
     u^2+x\right)\right){}^2}+\frac{b_3 u y}{b_3^2 u^2-a_2^2 \left(b_2
     u^2+x\right)}+\frac{x}{a_2}\\
    u&=z
\end{split}
\ee
while, inverting the second solution for $x,y$ we obtain
\be
\begin{split}
v&=\frac{a_2 y}{b_3^2 u^2-a_2^2 \left(b_2 u^2+x\right)}
  \\
  w&=\frac{b_3 u y}{b_3^2 u^2-a_2^2 \left(b_2 u^2+x\right)}-\frac{x}{a_2}\\
    u&=z
\end{split}
\ee

  These lead to the Weierstra\ss\, equation in Tate's form

\be
y^2+2\frac{b_3}{a_2}xyz\pm b_1a_2yz^3=x^3\pm \left(b_2-\frac{b_3^2}{a_2^2}\right)x^2z^2-b_0a_2^2xz^4-b_0a_2^2\left(b_2-\frac{b_3^2}{a_2^2}\right)z^6
\label{Weibi}
\ee
with the upper signs corresponding to the first case and the lower ones to the second solution.

Defining the functions
\be
\begin{split}
f&=b_1 b_3-a_2^2 b_0-\frac{b_2^2}{3}
  \\
g&=b_0 b_3^2+\frac{1}{12} a_2^2 \left(3 b_1^2-8 b_0 b_2\right)+\frac{2}{27} b_2^3-\frac{1}{3} b_1 b_3 b_2
\end{split}
\ee
we may also write down the compact Weierstra\ss\, form of the latter, which is just the form given in~(\ref{Weixyz}).

\section{ Constraints on Gauge Group Structure of the effective model}

After this short review we proceed with the investigation of the obtained Weierstra\ss\, form. The main point we wish to stress
 is that in the specific form given above,  the coefficients satisfy certain relations and therefore
are strongly constrained.   In this work  we restrict our  analysis to Weiestrass equation given by the original Tate's
algorithm~\cite{Tate1975,Bershadsky:1996nh}~\footnote{A generalisation of these results can be found in~\cite{Katz:2011qp}.}.
 Since the specific type of the non-abelian singularity depends on  the form of these
coefficients,   these aforementioned relations are expected to impose restrictions on the gauge group of the
effective theory. However, before abandoning the simple Tate algorithm, it is worth considering whether there are viable 
GUT symmetries left over to accommodate the Standard Model gauge group.
 To see this, we should  compare (\ref{Weibi})  with the general Tate form given by
 \be
 y^2+\alpha_1 xyz+\alpha_3yz^3=x^3+\alpha_2x^2z^2+\alpha_4 xz^4+\alpha_6z^6
 \ee
Comparing the two equations, we can extract the relations
\be
\label{TateCoef}
\begin{split}
\alpha_1&=\pm 2\frac{b_3}{a_2}\\
\alpha_2&=b_2-\frac{b_3^2}{a_2^2}\\
\alpha_3&=\pm b_1a_2\\
\alpha_4&=-b_0a_2^2\\
\alpha_6&=-\left(b_2-\frac{b_3^2}{a_2^2}\right)\,b_0a_2^2
\end{split}
\ee
Inspecting these equations, we can easily observe that the following relation holds among the coefficients
\be
\alpha_6=\alpha_2\alpha_4\label{a246}
\ee
Notice now that each of the coefficients can be represented locally by an expansion in the `normal' coordinate
$\xi$
\[ \alpha_n(\xi) = \alpha_{n,0}+\alpha_{n,1}\xi+\cdots\]
As is well known, the type of the geometric singularity associated to the non-abelian gauge group is determined
by the vanishing order of the coefficients $\alpha_n(\xi)$ with respect to $\xi$. For the most common non-abelian
symmetries these data are summarised in Table~\ref{TateTable}.

\begin{table}[!t]
\centering
\renewcommand{\arraystretch}{1.2}
\begin{tabular}{|c|c|c|c|c|c|c|c|}
\hline
{\bf Group} & ${  \alpha_1 }$& ${  \alpha_2 }$& ${  \alpha_3} $& ${  \alpha_4 }$& ${ \alpha_6} $& ${ \Delta}$&Type \\
\hline
${ SU(2n)}$ &0&1&$n$&$n$&$2n$&$2n$&  $I_{2n}^s$ \\
${ SU(2n+1)}$&0&1&$n$&$n+1$&$2n+1$&$2n+1$ & $I_{2n+1}^s$  \\
${ SO(10)}$&1&1&$2$&$3$&$5$&$7$ & $I_1^{*s}$  \\
${{\cal E}_6}$&1&2&$2$&$3$&$5$&$8$ & $IV^{*s}$  \\
${ {\cal E}_7}$&1&2&$3$&$3$&$5$&$9$ & $III^{*s}$  \\
${ {\cal E}_8}$&1&2&$3$&$4$&$5$&$10$ & $II^{s}$ \\
 \hline
\end{tabular}
 \caption{ Tate's algorithm for the most common non-abelian groups~\cite{Tate1975,Bershadsky:1996nh}.
  Table shows the {gauge group}, the order of vanishing of the coefficients ${ \alpha_k\sim a_{k,n}\xi^{n}}$,
   the discriminant $\Delta$
   and the corresponding  singularity type.\label{TateTable} }
\end{table}

We can examine now whether a relation of the form (\ref{a246}) can be fulfilled.

$\bullet$  From the first row
of the Table we can read off the relations of the coefficients for the $SU(2n)$ case. Indeed,
the vanishing order of $a_2$ is one, thus we may write $a_2=a_{2,1}\xi$, meaning that $a_{2,1}$ has
a constant part plus possible $\xi$-dependent terms. Similarly, in the same notation
we write $a_{4}=a_{4,n}\xi^n$ and $a_{6}=a_{6,2n}\xi^{2n}$. Hence,
\[ \alpha_2 \alpha_4 \propto \alpha_{2,1}\alpha_{4,n}\xi^{n+1}, \; \alpha_6\propto \alpha_{6,2n}\xi^{2n}\]
therefore the equation    $a_2a_4=a_6$ now reads
\[ \alpha_{2,1}\alpha_{4,n}\xi^{n+1}=\alpha_{6,2n}\xi^{2n}\; \Rightarrow\; n=1  \]
i.e., it is satisfied for  $n=1 $, corresponding to the $SU(2)$ group.

$\bullet$   For the $SU(2n+1)$ groups we have
\[ \alpha_2 \alpha_4 \propto \alpha_{2,1}\alpha_{4,n+1}\xi^{n+1}, \; \alpha_6\propto \alpha_{6,2n+1}\xi^{2n+1}\]
therefore the equation yields
\[ \alpha_{2,1}\alpha_{4,n+1}\xi^{n+2}=\alpha_{6,2n+1}\xi^{2n+1}\; \Rightarrow\; n=1  \]
which is satisfied for $n=1$ implying an $SU(3)$ group.

The above analysis shows that, in the context of Tate's form  for the $P_{(1,1,2)}$ case and the simple
mapping to Weierstra\ss\, model $P_{(1,2,3)}$ described in section 2, the only groups compatible with
the constraints of one additional rational section are $SU(3)$ and $SU(2)$. Extending our investigation  
to $SO(n)$ singularities, we infer that, 
if we restrict to the lower bounds on the vanishing orders of the
coefficients $\alpha_n(\xi)$ in Tate's algorithm,  the most common  GUT groups such as $SU(5)$ and $SO(10)$
are not accommodated.    To resolve this issue  a more detailed treatment is
required and a non-minimal version of the coefficients should be sought to  meet these conditions.
In fact, such GUT models can appear within the so called `top' constructions of 
toric geometry, which have been  studied in~\cite{Cvetic:2012xn}-\cite{Krippendorf:2014xba}.  Recently, the implementation 
of the latter  technique  was shown to  give rise to 
explicit constructions of various codimension one singularities.   However, we stress in this note that the familiar local 
Tate's forms are not completely excluded. Indeed,  repeating the analysis for the exceptional groups,  we will
 find out immediately, that the required criteria are fulfilled by two of them.

$\bullet$   For ${\cal E}_6$ we have
\[ \alpha_2 \alpha_4 \propto \alpha_{2,2}\alpha_{4,3}\xi^{5}, \; \alpha_6\propto \alpha_{6,5}\xi^{5}\]
i.e, the $\xi$ powers match and therefore we only need to impose the
equality constraint
\[\alpha_{2,2}\alpha_{4,3}=\alpha_{6,5}\]
 Once
this condition is satisfied, we also need to check the remaining coefficients constrained
by equations~(\ref{TateCoef}). To  investigate these relations, we  express all coefficients in terms
of  $a_2$. Assuming that the latter is given in terms of an unspecified power of the coordinate, $a_2\propto \xi^n$,
we find that a consistent solution exists in accordance with
\be
 b_0=-\alpha_{43}\xi^{3-2n},\; b_1=\alpha_{32}\xi^{2-n},\; b_2=(a_{22}+a_{11}^2/2)\xi^2,\; b_3=(a_{11}/2)\xi^{n+1}\label{b0123}
 \ee
 Requiring the  $b_0$ coefficient to be a positive power in $\xi$ we see that this  leaves two possibilities for the integer $n$, namely $n=0,1$.

\begin{table}[!t]
\centering
\renewcommand{\arraystretch}{1.2}
\begin{tabular}{|cccccc|}
\hline
Group& ${ a_2} $  & ${  b_0 }$& ${  b_1 }$& ${ b_2} $& ${ b_3 }$ \\
 \hline
 ${\cal E}_6$&1 &1&1&2&2\\
     &0 &3&1&2&1\\
  \hline
   ${\cal E}_7$&1 &1&2&2&2\\
       &0 &3&3&2&1\\
          \hline
\end{tabular}
 \caption{ The  vanishing order of the coefficients ${ b_k\sim b_{k,n}\xi^{n}}$,
of eq.~(\ref{hysus})  for the ${\cal E}_6$  and ${\cal E}_7 $ models\label{Tat} }
\end{table}
Substituting (\ref{b0123}) into the equations (\ref{TateCoef}) we find
\[ \alpha_{1}= \alpha_{11}\xi,\;  \alpha_{2}= \alpha_{2}\xi^2, \alpha_{3}= \alpha_{32}\xi^2,
 \alpha_{4}= \alpha_{43}\xi^3, \alpha_{6}= \alpha_{65}\xi^5\]
As can be checked in Table~\ref{Tat} this is just the requirement to obtain an ${\cal E}_6$ singularity.
We   compute the discriminant to find
\[ \Delta = -27 \alpha_{32}^4\xi^8+{\cal O}(\xi^9)\]
which, as expected has vanishing order 8.

$\bullet$ Repeating the analysis of the ${\cal E}_7$ case, we end up with the conditions on $b_i$'s
listed in the corresponding rows of Table~\ref{Tat}. Here, compared to the
 previous case,  we require also the vanishing of the
coefficient $\alpha_{32}$ so that $\alpha_3 = \alpha_{3,3}\xi^3$. It is also straightforward to see that
 $\Delta \propto \xi^9$ in accordance with Table~\ref{TateTable}.
Finally, notice that  for the ${\cal E}_8$ case, the condition $a_2a_4=a_6$ cannot be fulfilled.

\section{${\cal E}_6\times U(1)$ }

From the previous analysis, we have seen that in the presence of an additional rational section which
is associated to an extra $U(1)$ symmetry -as long as the minimal requirements on $\alpha_n$ of
Table~\ref{TateTable} are implemented-, the available non-abelian groups compatible with the
restrictions are $SU(3), SU(2)$ and the ${\cal E}_6$ and  ${\cal E}_7$.
 From these, only the
exceptional groups  are adequate to include the complete gauge symmetry of the SM.

The ${\cal E}_6$ model  has been extensively analysed in the literature. In the present context
the corresponding effective  model is based on the extended gauge group
 \[G_{GUT}={\cal E}_6\times U(1)  \]

 In the resulting effective theory all available matter is included in $78$  and $27$ representations.
 We can reduce the gauge symmetry down to the Standard Model using appropriate $U(1$) fluxes. We can reach
 the properties of the representations by successive decompositions of the ${\cal E}_6$ representations. The
 decomposition  ${\cal E}_6\to SO(10)\times U(1)_y$ gives
 \ba
78&\to &45_0+16_{-3}+\overline{16}_3+1_0\nn\\
27&\to & 16_1+\overline{10}_{-2}+1_4\nn
\ea
Under $SO(10)\to SU(5)\times U(1)_x$ the non-trivial representations obtain the following quantum numbers
\ba
45_0 &\rightarrow & 24_{(0,0)} + 10_{(4,0)} + \overline{10}_{(-4,0)} + 1_{(0,0)}\\
16_{-3} &\rightarrow & 10_{(-1,-3)} + \overline{5}_{(3,-3)} + 1_{(-5,-3)}\\
{\overline{10}}_{-2} &\rightarrow &{5}_{(2,-2)} + \overline{5}_{(-2,-2)}
\ea
and analogously for the other representations, while the $SU(5)$ singlet emerging from 27  is  $ 1_{(0,4)}$.

Observe that $10,5$'s of $SU(5)$  emerge from $27$ as well as $78$ so it is possible to accommodate families
in both.  In the simplest scenario the third family fermions and the Higgs fields reside  in  $27_{q},\, 27_{q'}$.
 To write down superpotential terms of the effective model, we need the  charges $q,q'$ under the
Mordell-Weil $U(1)$.  This computation is rather involved and goes beyond the scope of this short  note. However,
in analogy with $SU(5)$  models, we might expect a solution where the allowed charges are multiples of
 $1/3$ so that a tree level coupling of the form could be allowed
\ba
 27_{\frac 13}\, 27_{\frac 13}\, 27_{-\frac 23}&\ra& 10_M \,10_M \,5_{h_u}+
 10_M \,\bar 5_M \,\bar 5_{h_d}\to m_t,\, m_b
 \ea
As indicated, this is suitable to derive the top and bottom quark entries, while higher order terms involving
powers of the {78}-representation can give higher order contributions to the fermion masses of
the lighter generations
 \ba
({ 78}\,+\,{ 78}^2) 27_{\frac 13}\, 27_{\frac 13}\, 27_{-\frac 23}&\ra& m_{u_{ij}},\;  m_{d_{ij}}
\ea
A detailed analysis of the ${\cal E}_6$ F-theory models is beyond the scope of this note and can be found
in~\cite{Callaghan:2011jj}.

\section{Conclusions}

In this note we investigated constraints on GUTs in F-theory compactifications
with  an extra  rational section  which corresponds to an additional
abelian factor in the gauge group of the final  effective theory model.
 Elliptic fibrations with two sections can be represented
by a quartic polynomial of definite form  written in terms of three homogeneous  coordinates
in the ambient space $P_{(1,1,2)}$. 
Converting the quadratic equation to a local Tate from we find that the Tate coefficients 
are subject to constraints which restrict the number of non-abelian gauge groups that can be 
realized in the local Tate form. 
 Models emerging in this context which can accommodate the Standard Model
 gauge symmetry are based on ${\cal E}_6\times U(1)$ and ${\cal E}_7\times U(1)$.
  We discuss briefly the salient features of the ${\cal E}_6\times U(1)$  case.

\newpage

\section*{Acknowledgements}

This work is supported in part by the European Commission under the ERC Advanced Grant 226371.
This research has been
co-financed by the European Union (European Social Fund - ESF) and
Greek national funds through the Operational Program "Education and
Lifelong Learning" of the National Strategic Reference Framework
(NSRF) - Research Funding Program: "ARISTEIA". Investing in the
society of knowledge through the European Social Fund.

\end{document}